\documentstyle[useAMS,graphics,psfig]{mn2e}
\def\zabs{$z_{\rm abs}$}

\def\lya{Ly$\alpha$ }

\def\h2{H$_2$}
\def\hi{H{\sc i}~}

\def\kms{km~s$^{-1}$}
\begin{document}
%
\title[Dust Depletion in Damped Lyman-$\alpha$ systems]{Molecular 
hydrogen at \zabs~=~1.973 toward Q~0013$-$004:
Dust depletion pattern in damped Lyman-$\alpha$ systems
\thanks{Based on observations carried out at the European Southern 
Observatory (ESO) under prog. ID No. 66.A-0624 and 267.A-5714 with UVES 
the echelle spectrograph mounted on the Very Large Telescope (VLT) at 
Cerro Paranal Observatory in Chile.}
}
%
%
\author[Patrick Petitjean, R. Srianand, \& Cedric Ledoux]{
Patrick Petitjean$^{1,2}$, R. Srianand$^{3}$, \& C\'edric Ledoux$^{4}$\\ 
   ${}^1$ Institut d'Astrophysique de Paris -- CNRS, 98bis Boulevard 
   Arago, F-75014 Paris, France - email: petitjean@iap.fr\\
   ${}^2$ LERMA, Observatoire de Paris, 61, avenue de l'Observatoire, 
   F-75014, Paris, France\\
   ${}^3$ IUCAA, Post Bag 4, Ganeshkhind, Pune 411 007, India -
   email: anand@iucaa.ernet.in\\
   ${}^4$ European Southern Observatory, Karl-Schwarzschild Stra$\ss$e 2,
   D-85748 Garching bei M\"unchen, Germany - email: cledoux@eso.org
}
\date{Typeset \today ; Received / Accepted}
\maketitle
\begin{abstract}
We study the dust depletion pattern in eight well separated components of the 
$z_{\rm abs}$~=~1.973, log~$N$(H~{\sc i})~=~20.83, damped Lyman-$\alpha$
system toward Q~0013$-$004, four of which have detectable H$_2$ absorption. 
The apparent correlation between the abundance ratios 
[Fe/S] and [Si/S] in the components indicates that the abundance pattern is 
indeed due to dust-depletion. In particular, we find evidence for depletion 
similar to what is observed in cold gas of the Galactic disk 
([Fe/Zn]~=~$-$1.59, [Fe/S]~=~$-$1.74, [Zn/S]~=~$-$0.15, [Si/S]~=~$-$0.85)
in one of the weakest components  in which molecular hydrogen is detected 
with log~$N$(H$_2$)~$\sim$~16.5. This is the first time such a large depletion 
is seen in a DLA system.  Extinction due to this component is negligible 
owing to small total \hi column density, log~$N$(\hi)~$\leq$~19.4.
This observation supports the possibility that current samples of DLA 
systems might be biased against the presence of cold and dusty gas along the 
line of sight.
\par\noindent
The overall metallicities of this peculiar DLA system in which O~{\sc i} and 
C~{\sc ii} are spread over $\sim$1050~km~s$^{-1}$ are [P/H]~=~$-$0.64, 
[Zn/H]~=~$-$0.74 and [S/H]~=~$-$0.82 relative to solar. The sub-DLA system 
at $z_{\rm abs}$~=~1.96753 has [P/H]~$>$~0.06, [Zn/H]~$>$~$-$0.02 and 
[S/H]~$>$~$-$0.18. The overall molecular fraction is in the range 
$-$2.7~$<$~log~$f$~$<$~$-$0.6, which is the highest 
value found for DLA systems. H$_2$ is detected in four components at $-$615, 
$-$480, 0 and 85~km~s$^{-1}$ relative to the strongest component at 
$z_{\rm abs}$~=~1.97296. CO is not detected 
(log~$N$(CO)/$N$(H~{\sc i})~$<$~$-$8) and HD could be present at 
$z_{\rm abs}$~=~1.97380.
\par\noindent
We show that the presence of \h2 is closely related to the physical 
conditions of the gas: high particle density together with low temperature. 
The observed excitation of high $J$ H$_2$ levels and the molecular fraction 
show large variations from one component to the other suggesting that the 
UV radiation field is highly inhomogeneous throughout the system.
Gas pressure, estimated from C~{\sc i} absorptions, is larger than what is 
observed in the ISM of our Galaxy. This, together with the complex kinematics,
suggests that part of the gas is subject to high compression due to either 
collapse, merging and/or supernovae explosions. This is probably a consequence 
of intense star-formation activity in the vicinity of the absorbing gas.
\end{abstract}
\section{Introduction}
The amount of dust present at high redshift has important consequences on the 
physics of the gas.
In addition, dust directly affects our view of the high redshift universe 
through extinction. Therefore, the presence of dust in damped Lyman-$\alpha$
(hereafter DLA) systems, that contain most of the neutral hydrogen in the 
universe, can have significant consequences. 
Although the presence of dust in DLA systems has been 
claimed very early (Pei et al. 1991), the issue has remained controversial. 
Indeed, Lu et al. (1996) have questioned the idea that the 
overabundance of Zn compared to Cr or Fe observed in DLA systems (e.g. Pettini 
et al. 1997) is due to selective depletion onto dust-grains and have argued
that the overall abundance pattern observed in DLA systems is indicative 
of Type II supernovae enrichment instead. In recent years several studies 
have shown that both effects, dust-depletion and peculiar nucleosynthesis 
history, should be invoked to explain the abundance pattern (Vladilo 1998, 
Prochaska \& Wolfe 1999, Ledoux et al. 2001a). However, the lack of statistics 
and the wide variety of objects that can give rise to DLA systems, namely dwarf 
galaxies (Centuri\'on et al. 2000), large disks (Prochaska \& Wolfe 1997, 
Hou et al. 2001), galactic building blobs (Haehnelt et al. 1998, Ledoux et 
al. 1998) etc., with, for each of these objects, its own history, prevent 
us to have a clear picture of the nature of DLA systems. Nevertheless, all 
studies conclude that the dust content of DLA systems is small. However, it is 
possible that the current sample of DLA systems is biased against 
high-metallicity and dusty systems. Indeed, Boiss\'e et al. (1998) have 
noticed that there is a lack of systems with large $N$(H~{\sc i}) and large 
metallicity. Very recent investigation of an homogeneous sample of radio-selected quasars shows that 
the dust-induced bias cannot lead to underestimate the H~{\sc i} mass in 
DLA systems by a large factor (Ellison et al. 2001). However even a 
factor of two could change our understanding of DLA systems.
\par\noindent
An obvious way to search for DLA systems with large amount of dust is to select 
those where molecules are detected as these molecules form predominantly
at the surface of dust grains. However, it may not be so simple as it has been shown that 
the presence of H$_2$ is not only related to the dust-to-metal ratio but is 
mostly dependent on the physical conditions of the gas. First of 
all, H$_2$ is detected when the particle density is large (Petitjean et al. 
2000, Ledoux et al. 2001b). In any case, the system at $z_{\rm abs}$~=~1.973 
toward Q~0013$-$004 is a good target as very strong molecular absorption
lines have been identified by Ge \& Bechtold (1997, see also Ge et al. 1997). 
\section{Observations}
The~Ultra-violet and Visible Echelle Spectrograph (D'Odorico et al. 2000) 
mounted on the ESO Kueyen 8.2~m telescope at the Paranal observatory has 
been used on October 21$-$23, 2000, to obtain high-spectral resolution spectra 
of Q~0013$-$004. Additional observations were performed in service mode on 
September 10$-$11 and 13$-$18 2001. The slit width was 1~arcsec (the seeing $FWHM$ 
was most of the time better than 0.8~arcsec) and the CCDs were binned 
2$\times$2. The resulting resolution was $\sim$45000. The total exposure time 
$\sim$22~hours was split into 1~h exposures. Two dichroic settings were used 
to cover the complete wavelength range. The data were reduced using MIDAS, 
the ESO data reduction package, and the UVES pipeline in an 
interactive mode. The main characteristics of the pipeline is to perform a 
precise inter-order background subtraction, especially for master 
flat-fields, and an optimal extraction of the object signal rejecting cosmic 
ray impacts and subtracting the sky at the same time. The data reduction is 
checked step by step. Wavelengths were corrected to vacuum-heliocentric 
values and individual 1D spectra were combined together. The resulting S/N 
ratio per pixel in the final spectrum is of the order of 35 at 
$\sim$3500~\AA~ and 80 at $\sim$5000~\AA.
\section {The DLA system at $z_{\rm abs}$~=~1.973}
\subsection{Molecular hydrogen}
\begin{figure*}
\centerline{\vbox{
\psfig{figure=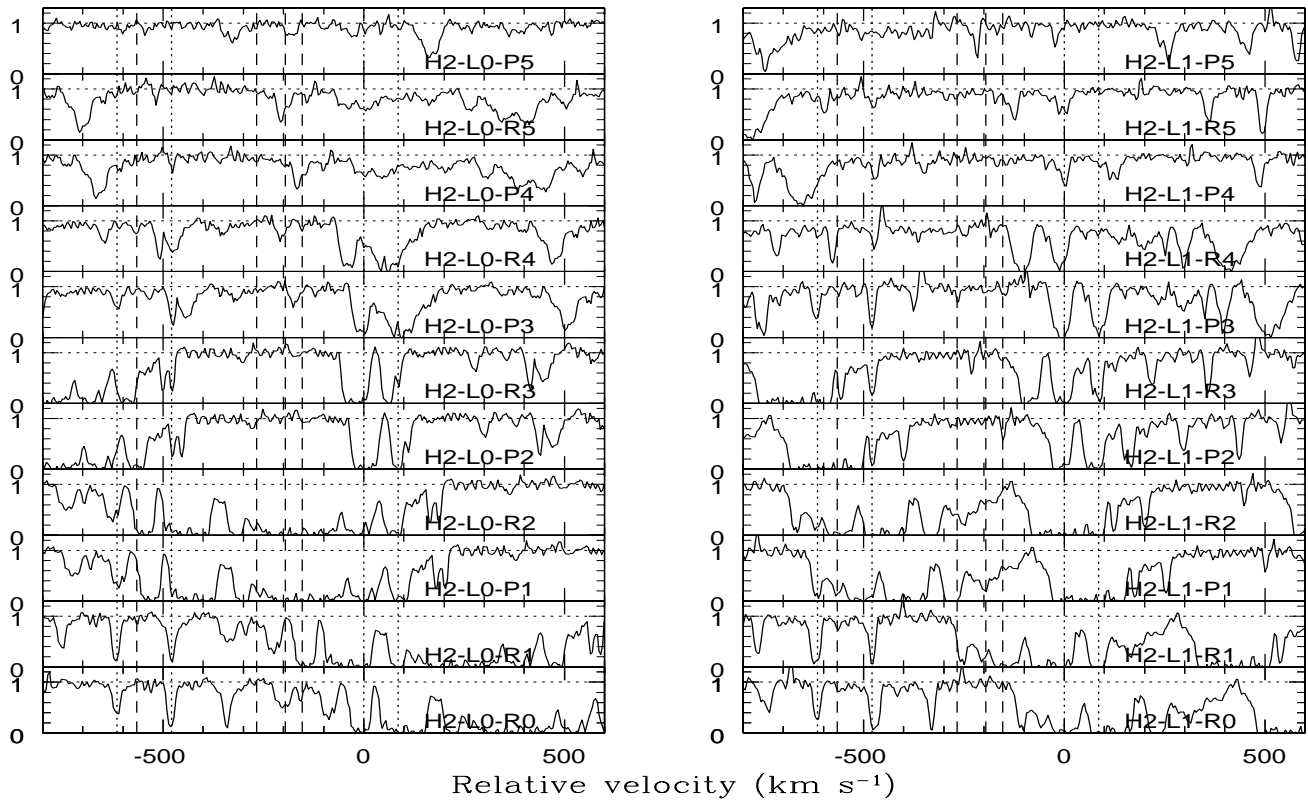,height=11.5cm,width=20.cm,angle=270}
}}
\caption[]{\h2 absorption profiles for transitions from different levels, 
${\large v} = 0-0$ (left panels) and $1-0$ (right panels), given on a velocity 
scale with origin at \zabs = 1.97296. The vertical dotted lines 
indicate the positions of the four \h2 components discussed 
in the text. The location of components with strong O~{\sc i}, Si~{\sc ii},
Mg~{\sc ii} and S~{\sc ii} absorption lines but without detectable 
absorption due to \h2 is indicated by dashed vertical lines (see also Fig.~2).
}
\label{h2fig}
\end{figure*}
\begin{table}
\caption{Fit results for \h2 lines}
\begin{tabular}{llccr}
\hline
\multicolumn{1}{c}{$z_{\rm abs}$} &\multicolumn {1}{c}{Level}&
\multicolumn{1}{c}{log $N$}& \multicolumn {1}{c}{$n$$_{\rm lines}$}&
\multicolumn{1}{c}{$T_{\rm ex}$}\\
\multicolumn{1}{c}{ }&\multicolumn{1}{c}{ }&
\multicolumn{1}{c}{(cm$^{-2}$)}&\multicolumn{1}{c}{ }&\multicolumn{1}{c}{(K)}
\\
\hline
1.96685$^{\rm a}$ & J = 0 &2.35$\pm0.25\times10^{15}$&3&\multicolumn{1}{c}{...}\\
         & J = 1 & 1.21$\pm0.19\times10^{16}$&2&300$_{-96}^{+276}$\\
         & J = 2 & 6.09$\pm1.10\times10^{15}$&1&776$_{-244}^{+584}$\\
	 & J = 3 & 3.39$\pm0.39\times10^{15}$&2&
                     \multicolumn{1}{c}{382$_{-29}^{+35}$} \\
	 & J = 4 & $\le2.5\times10^{14}$&\multicolumn{1}{c}{...}&\multicolumn{1}{c}{...} \\
1.96822$^{\rm b}$  & J = 0 & 1.05$\pm0.08\times10^{16}$:&1 &\multicolumn{1}{c}{...} \\
         & J = 1 & 9.03$\pm1.61\times10^{15}$&2 & 73$_{-8}^{+7}$\\
         & J = 2 & 9.65$\pm1.30\times10^{15}$&1 & 302$_{-30}^{+33}$\\
         & J = 3 & 4.18$\pm0.43\times10^{15}$&3 & 258$_{-12}^{+6}$\\
         & J = 4 & 8.58$\pm0.68\times10^{14}$&3 & 363$_{-12}^{+10}$\\
         & J = 5 & 3.57$\pm0.42\times10^{14}$&5 & 372$_{-11}^{+9}$\\
1.97296$^{\rm c}$  & J = 0 & 5.5$\times10^{16}$--1.0$\times10^{19}$&\multicolumn{1}{c}{...}&\multicolumn{1}{c}{...}   \\
         & J = 1 & 1.4$\times10^{17}$--2.0$\times10^{19}$&\multicolumn{1}{c}{...}&\multicolumn{1}{c}{...}  \\
         & J = 2 & 7.5$\times10^{16}$--2.0$\times10^{19}$&\multicolumn{1}{c}{...}&\multicolumn{1}{c}{...}  \\
         & J = 3 & 4.3$\times10^{16}$--1.0$\times10^{19}$& \multicolumn{1}{c}{...}&\multicolumn{1}{c}{...} \\
         & J = 4 & 1.46$\pm0.56\times10^{15}$&\multicolumn{1}{c}{3}&\multicolumn{1}{c}{...} \\
         & J = 5 & 1.30$\pm0.14\times10^{15}$&\multicolumn{1}{c}{4}&\multicolumn{1}{c}{...} \\
1.97380$^{\rm d}$  & J = 0 & 5.0$\times10^{16}$--6.0$\times10^{18}$&...&\multicolumn{1}{c}{...} \\
         & J = 1 & 1.0$\times10^{17}$--3.0$\times10^{19}$&...&\multicolumn{1}{c}{...} \\
         & J = 2 & 5.0$\times10^{16}$--3.0$\times10^{18}$&...&\multicolumn{1}{c}{...} \\
         & J = 3 & 6.42$\pm0.91\times10^{15}$&1&\multicolumn{1}{c}{...}\\ 
	 & J = 4 & $\le5\times10^{14}$&...&\multicolumn{1}{c}{...} \\
\hline
\multicolumn{5}{l}{$^{\rm a}$ $b$~=~8.89$\pm0.46$~km~s$^{-1}$;
$^{\rm b}$ $b$~=~7.07$\pm0.33
$~km~s$^{-1}$}\\
\multicolumn{5}{l}{$^{\rm c}$ $b$~=~18.87$\pm2.74$~km~s$^{-1}$;
$^{\rm d}$ $b$~=~11.26$\pm0.93
$~km~s$^{-1}$}
\end{tabular}
\label{taba}
\end{table}
The presence of \h2 in the \zabs~=~1.973 system toward Q~0013$-004$ was first 
reported by Ge \& Bechtold (1997). Based on an intermediate resolution spectrum
and a single component curve of growth analysis, they derived a very high 
value of the 
molecular fraction, $f$~=~2 $N$(\h2)/[$N$(H~{\sc i})+2$N$(\h2)] = 
$0.22\pm0.05$. In the UVES data, \h2 is detected in four distinct main 
components spread over $\sim$700~km~s$^{-1}$, at \zabs = 1.96685 ($\sim-615 $ 
\kms), 1.96822 ($\sim-$480 \kms), 1.97296 (0 \kms) and 1.97380 ($\sim+85 $ 
\kms), marked with vertical dotted lines in Fig.~\ref{h2fig}. For simplicity 
we name these components {\bf a}, {\bf b}, {\bf c} and {\bf d} respectively.
Four additional strong metal components at \zabs = 1.96729 ($\sim-572 $ \kms), 
1.97023 ($\sim-275 $ \kms), 1.97098 ($\sim-200 $ \kms) and 1.97138 
($\sim-159 $ \kms), with highly saturated  O~{\sc i}, Si~{\sc ii} and 
Mg~{\sc ii} lines, {\sl but clearly identified by the weaker S~{\sc ii} 
lines}, are marked with vertical dashed lines in Fig.~\ref{h2fig} (see also 
Fig.~\ref{metpro} and Fig.~4). These additional components, in which we 
measure a conservative upper limit, $N$(\h2)$\le10^{14}~{\rm cm}^{_2}$, 
probably have a non-negligible contribution to the total H~{\sc i} column 
density. 
\par\noindent
In the case of components {\bf c} and {\bf d} the absorption lines due to 
$J\le3$ rotational levels are strongly saturated and possibly blended with 
other lines. This implies that we can only derive a lower limit on the \h2 
column density from the condition that the lines must be saturated and an 
upper limit from the absence of damping wings. Results are given in 
Table~\ref{taba} together with the number of unblended lines, $n_{\rm lines}$, 
that are used to derive the column densities.
\par\noindent
Components {\bf a} and {\bf b} are narrow and mostly unblended. 
As the lines are moderately saturated it is possible to derive \h2 
column densities for various $J$ levels (see Table~\ref{taba}). The total 
\h2 column density in these components is 2.39$\pm0.22\times10^{16}$ and 
3.46$\pm0.40\times10^{16}~{\rm cm^{-2}}$ respectively. The excitation 
temperatures $T_{01}$~=~300$^{+276}_{-96}$ and 73$^{+7}_{-8}$~K
could reflect the kinetic temperature of the gas (e.g. Abgrall et al. 1992).
\par\noindent
In the case of component {\bf a} the excitation temperatures for the ortho 
and para states ($T_{02}=776_{-244}^{+584}$ K and $T_{13}=401_{-55}^{+62}$ K)
are different (although errors are large). This happens also in component
{\bf b}. Note however that the estimated column densities of $J=0$ and $J=2$ 
levels are based on a single unblended line. Moreover, $J = 0$ transitions 
from this latter component are blended with $J = 1$ transitions at 
\zabs=1.96685 
(see Fig.~\ref{h2fig}). The fact that $T_{02}$ and $T_{13}$ are different 
could indicate that excitation is not dominated by collisions and other 
processes such as UV pumping are important (see Srianand \& Petitjean 1998).
As emphasized by Srianand et al. (2000), populating the J~=~4 level after 
formation of a molecule on a dust grain is probably negligible in this gas. 
The photo-absorption rate in the Lyman and Werner bands is therefore
$\beta_o\le10^{-9}$ and 9$\times10^{-10}$~s$^{-1}$, respectively, for {\bf a} 
and {\bf b}, based on the $J=4$ column densities. This is consistent
with what is measured in the Galactic ISM.
\par\noindent
Absorption due to \h2 is detected up to $J=5$ in component {\bf c}. 
The lines from the $J$~=~4 and 5 levels are narrow and unsaturated and we used a 
single-component fit to derive the column densities. Neglecting the formation pumping 
to $J=4$ and using the lower limit on the $J=0$ column density, we obtain 
$\beta_o\le3.0\times10^{-10}{\rm s}^{-1}$. This is at least a factor 3 smaller than 
that derived for component {\bf b}.
\par\noindent
It is clear from Fig.~\ref{h2fig} that the $J\ge4$ lines are either 
absent or weak in component {\bf d}. We determine the column density for 
$J = 3$ from the H$_2$-L0-R3 line that is not saturated. 
Absence of absorption from high $J$ levels, despite strong saturation of the
transitions from $J\le2$ levels, suggests that the rate of UV pumping to high 
$J$ levels is low in this component. We derive  
$\beta_o\le1.1\times10^{-10} {\rm s}^{-1}$ which is at least a factor 8 
smaller than what is derived for component {\bf  b}.
\par\noindent
All this indicates that there are large fluctuations of the ambiant
UV flux inside the system.
\par\noindent
\subsection{Atomic hydrogen}
\begin{figure*}
\centerline{\vbox{
\psfig{figure=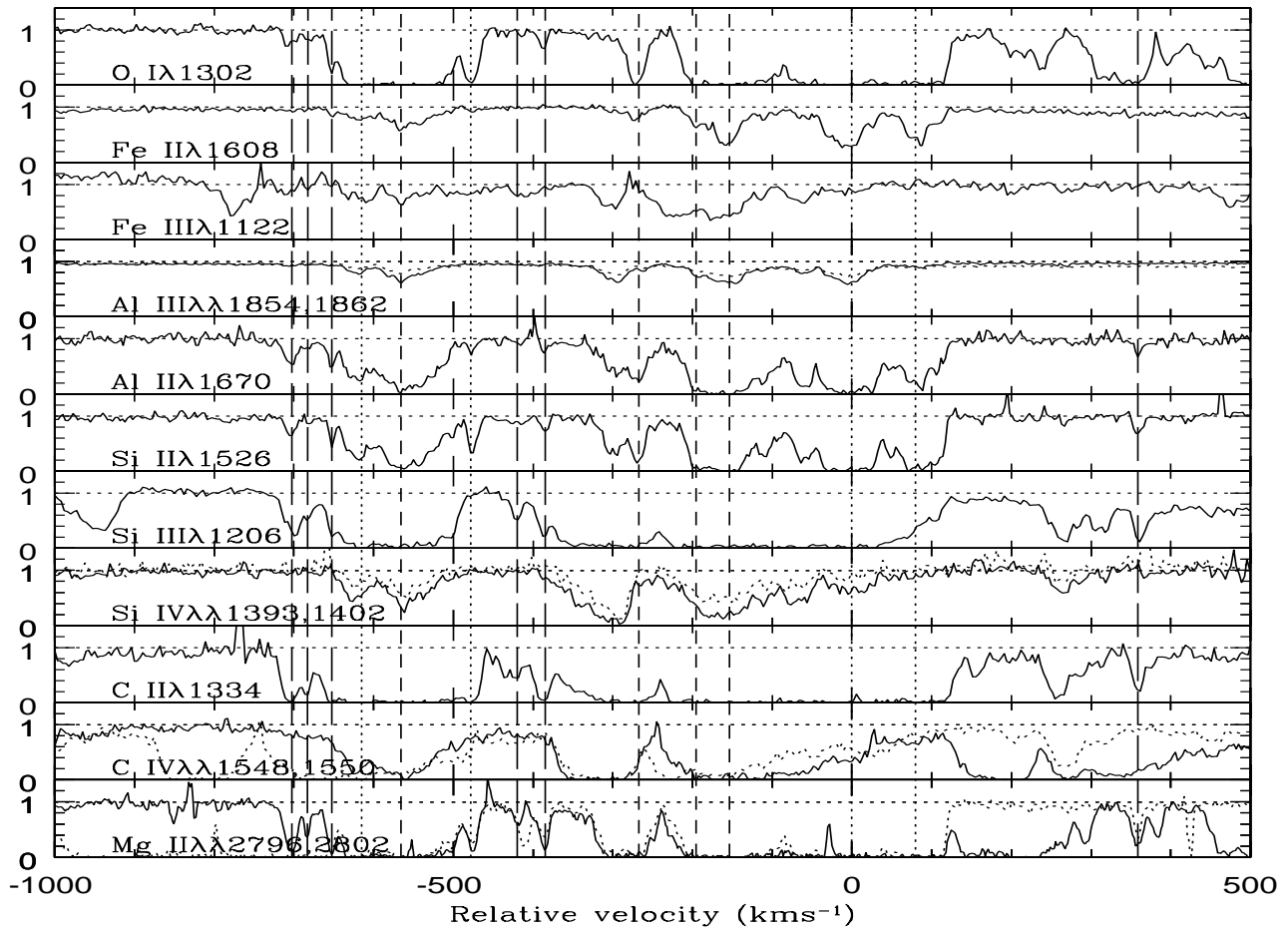,height=13.5cm,width=19.cm,angle=0}
}}
\caption[]{Metal line profiles on a velocity scale with origin at redshift 
$z_{\rm abs}$~=~1.97296. The vertical dotted and short-dashed lines are 
as in Fig.~\ref{h2fig}. The vertical long-dashed lines indicate narrow 
components seen only in low-ionization species. For doublets, the profile
of the weakest line is overplotted as a dashed line.
}
\label{metpro}
\end{figure*}
\begin{figure}
\centerline{\vbox{
\psfig{figure=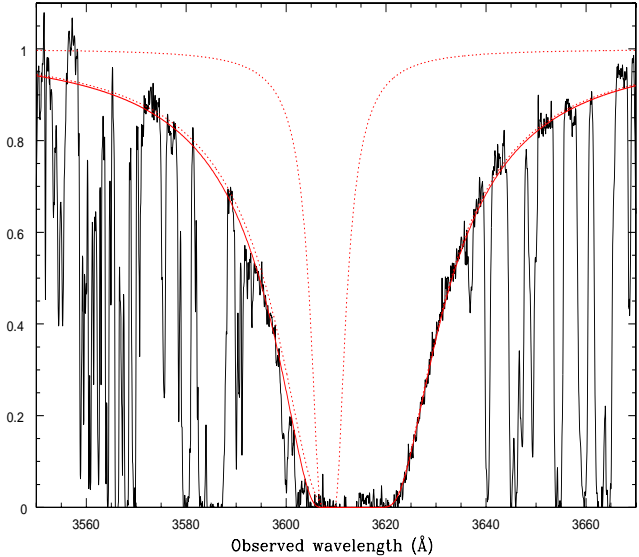,height=8.cm,width=9.5cm,angle=270}
}}
\caption[]{Fits to the damped \lya absorption profile at 
$z_{\rm abs}$~=~1.9733. Fits with (solid line) and without (dotted line) the 
maximum column density (log~$N$(H~{\sc i})~=~19.4) present in the H$_2$ 
components {\bf a} and {\bf b} at a mean redshift $z_{\rm abs}$~=~1.96753, 
compatible with the asymmetry of the profile.
}
\label{lyafit}
\end{figure}
The overall structure of the strong metal lines is illustrated in 
Fig.~\ref{metpro}. The location of the strongest components with and without 
detected H$_2$ molecules is indicated by, respectively, dotted and 
short-dashed lines as in Fig.~\ref{h2fig}. The long-dashed lines show 
components with absorption due to low-ionization species only. It can be seen 
that the low-ionization absorptions (from O~{\sc i}, C~{\sc ii}, Si~{\sc ii}, 
Mg~{\sc ii}) span about 1000~km~s$^{-1}$ and are structured in two main 
strong systems at $z_{\rm abs}$~$\sim$~1.9676 and 1.9733 separated by 
575~km~s$^{-1}$. Note that most of the Si~{\sc iv} and C~{\sc iv} absorption 
is located inbetween these two strong low-ionization systems.
\par\noindent
A Voigt profile centered at the mean position of the two strongest H$_2$ 
components ({\bf c} and {\bf d}), $z_{\rm abs}$~=~1.9733, gives a good fit 
to the red wing of the Lyman-$\alpha$ absorption with a total column density, 
$N$(H~{\sc i}) = 6.7$\times$10$^{20}$~cm$^{-2}$. This is consistent with
the value derived by Ge \& Bechtold (1997). There is however an excess of 
absorption in the blue wing that we attribute to H$_2$ components {\bf a} and 
{\bf b} that are blue-shifted by 615 and 480~km~s$^{-1}$ relative to 
the strongest H$_2$ component. 
We can therefore estimate $N$(H~{\sc i}) associated with these two blueshifted
components. In Fig.~\ref{lyafit}, we show the fits to the damped absorption 
with and without the additional absorption (log~$N$(H~{\sc i})~=~19.4) that 
can be accomodated at $z_{\rm abs}$~$\sim$~1.96753. This implies that less 
than 10\% of the \hi gas is associated with H$_2$ components {\bf a} and 
{\bf b}.
This gives a lower limit on the molecular fraction $f$~$>$~$2\times10^{-3}$ 
and $4\times10^{-3}$  for components {\bf a} and {\bf b} respectively. These 
limits are higher than what has been observed in other DLA systems 
(e.g. Petitjean et al. 2000). Such a large value for $f$ in a low \hi 
column density cloud is at odd with what is observed in the ISM of our 
Galaxy where log~$f$~$<$~$-$4 for log~$N$(H~{\sc i})~$<$~20 (Savage et al. 
1977). This implies high formation rate, and therefore large amount of dust, 
high density, low temperature, {\sl and} low destruction rate, therefore 
shielding from the UV field.  
\par\noindent
We derive for the whole system 
6$\times$10$^{17}$ $<$ log~$N$(\h2) $<$ 1$\times10^{20}$~cm$^{-2}$.  
Therefore, the mean fraction of molecular hydrogen is in the range 
0.002$-$0.25.  This is much larger than what is seen in other DLA systems 
(see Petitjean et al. 2000; Ledoux et al. 2001b). 
\subsection{Other molecules}
If the molecular fraction is close to the upper limit we derived
above then one can expect to detect other molecular species such as
CO and HD (see Varshalovich et al. 2001).
We do not detect molecular CO down to $w_{\rm obs}$($\lambda$1088)~$<$~7~m\AA.
This means $N$(CO)~$<$~2$\times$10$^{12}$~cm$^{-2}$ in components {\bf c} 
and {\bf d}, using the oscillator strength by Morton (1975). 
Therefore log~$N$(CO)/$N$(H~{\sc i})~$<$~$-$8.5. Note that
in our Galaxy, when log~$N$(H$_2$)~$>$~19, CO is readily detected with 
log~$N$(CO)~$\ge$~13 (Federman et al. 1980). 
\par\noindent
It is difficult to trace HD in this system because blending of lines
is severe for all bands. At $z_{\rm abs}$~=~1.97296 (component {\bf c}),
absorption features are present at the expected position of HD-L0-R0 
and HD-L1-R0 but they are stronger than a possible HD-L3-R0 absorption 
($w_{\rm obs}$~$\sim$~30~m\AA). Note that consistent features are
seen at $z_{\rm abs}$~=~1.97278, especially at the expected position
of HD-L3-R0 but HD-L0-R0 is not visible down to $w_{\rm obs}$~$<$~40~m\AA.
The large equivalent width limit is a consequence of the line falling
in the wing of another stronger absorption. At $z_{\rm abs}$~=~1.97380 
(corresponding to component {\bf d}), consistent absorption features are 
present at the expected position of HD-L0-R0, HD-L3-R0 and HD-L4-R0. 
HD-L1-R0 is possibly present but blended. The detection should be considered 
as tentative as HD-L0-R0 and HD-L3-R0 are affected by noise. The strongest 
constraint on this component is given by 
$w_{\rm obs}$(HD-L0-R0)~$\sim$~10~m\AA. Using $\lambda$~=~1105.845~\AA~ 
and $f$~=~7.6$\times$10$^{-4}$ we derive 
$N$(HD)~$\sim$~4$\times$10$^{14}$~cm$^{-2}$. Using the limits on 
\h2 indicated in Table~1, we derive
1.0$\times$10$^{-5}$~$<$~$N$(HD)/$N$(H$_2$)~$<$~2$\times$10$^{-3}$.
\par\noindent
\section{Metallicity and depletion}
\subsection{The system as a whole}
\begin{figure}
\centerline{\vbox{
\psfig{figure=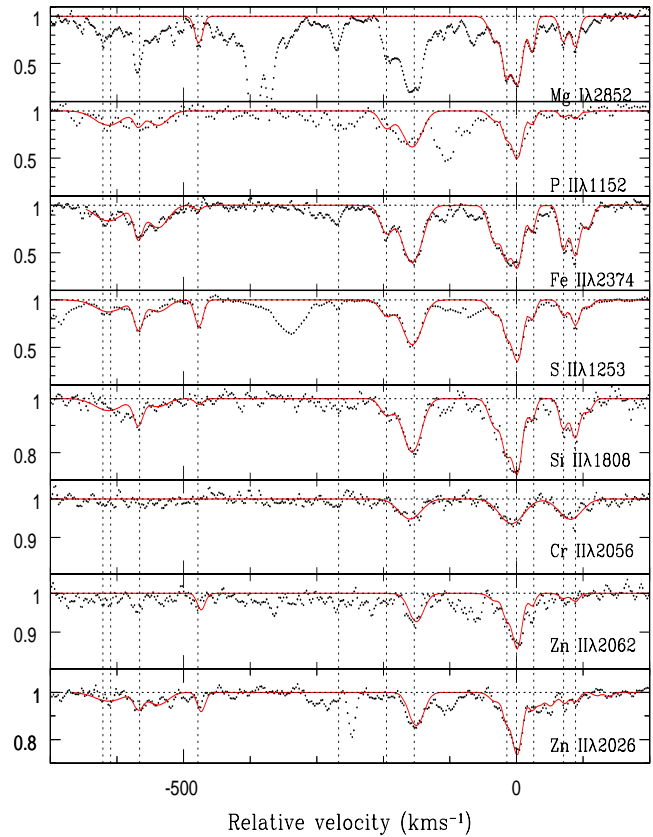,height=12.cm,width=9.5cm,angle=0}
}}
\caption[]{Profiles of various metal lines detected in the DLA system plotted 
on a velocity scale with origin at $z_{\rm abs}$~=~1.97296. Solid lines  
give the best Voigt profile fits. Vertical dotted lines indicate the 
positions of subcomponents.
}
\label{metfit1}
\end{figure}
\begin{figure}
\centerline{\vbox{
\psfig{figure=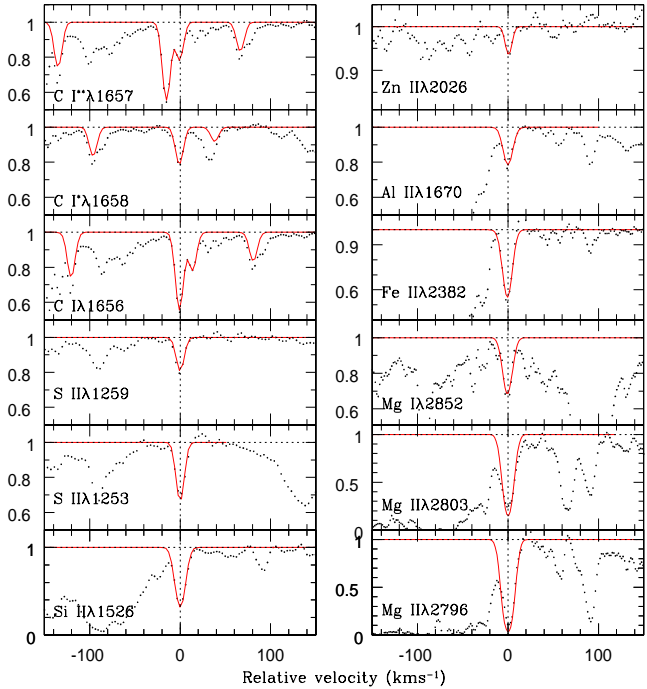,height=10.cm,width=9.5cm,angle=0}
}}
\caption[]{Profiles of various metal lines detected in the H$_2$ component at
\zabs = 1.96822 (component {\bf b}). Solid lines give the best Voigt profile 
fits.}
\label{metfit}
\end{figure}
\par\noindent
The absorption lines from heavy elements show a complex multiple component 
structure (see Fig.~\ref{metfit1}) and only a few of the strong velocity 
components have detectable associated \h2 absorption. 
\begin{table}
\caption{Total column densities and metallicities relative to solar}
\centerline{
\begin{tabular}{lccccc}
\hline
\multicolumn{1}{c}{Ion}&\multicolumn{1}{c}{$N$}&
\multicolumn{4}{c}{[X/S]$^{\rm a,b}$}\\
\multicolumn{1}{c}{ }&\multicolumn{1}{c}{(cm$^{-2}$)}&
\multicolumn{1}{c}{DLA}&\multicolumn{1}{c}{Cold}&
\multicolumn{1}{c}{Warm}&\multicolumn{1}{c}{Halo}\\
\hline
\multicolumn{6}{c}{DLA at $z_{\rm abs}$~=~1.9733$^{\rm c}$}\\
H~{\sc i}    &  6.70$\pm0.70\times10^{20}$&......&.....&......&......\\
Si~{\sc ii}  &  2.70$\pm0.20\times10^{15}$&$-0.13$&$-1.3$&$-0.4$&$-0.3$\\
S~{\sc ii}   &  1.90$\pm0.10\times10^{15}$&$-0.00$&....&$+0.0$&$+0.0$\\
P~{\sc ii}   &  5.71$\pm0.77\times10^{13}$&$+0.18$&$-0.5$&$-0.2$&$-0.1$\\
Zn~{\sc ii}  &  5.47$\pm0.50\times10^{12}$&$+0.08$&$-0.4$&$-0.2$&$-0.1$\\
Cr~{\sc ii}  &  9.86$\pm0.30\times10^{12}$&$-0.70$&$-2.1$&$-1.2$&$-0.6$\\
Fe~{\sc ii}  &  6.41$\pm0.46\times10^{14}$&$-0.71$&$-2.2$&$-1.4$&$-0.6$\\
Ni~{\sc ii}  &  1.44$\pm0.60\times10^{13}$&$-1.10$&$-2.2$&$-1.4$&$-0.6$\\
Mn~{\sc ii}  &  2.31$\pm0.50\times10^{12}$&$-1.18$&$-1.5$&$-1.0$&$-0.7$\\
\hline
\multicolumn{6}{c}{Sub-DLA at $z_{\rm abs}$~=~1.96753$^{\rm d}$}\\
H~{\sc i}    &  $<$2.70$\pm0.20\times10^{19}$&.... \\
Si~{\sc ii}  &  2.65$\pm0.31\times10^{14}$&$-$0.38& \\
S~{\sc ii}   &  3.32$\pm0.29\times10^{14}$&$-0.00$&\\
P~{\sc ii}   &  1.16$\pm0.17\times10^{13}$&$+0.24$&\\
Zn~{\sc ii}  &  1.16$\pm0.12\times10^{12}$&$+0.16$&\\
Fe~{\sc ii}  &  5.43$\pm0.35\times10^{13}$&$-1.03$&\\
\hline
\multicolumn{6}{l}{$^{\rm a}$[X/H]~=~log[$Z$(X)]$-$log[$Z_{\odot}$(X)]}\\
\multicolumn{6}{l}{$^{\rm b}$Galactic values are from Welty et al. (1999)}\\
\multicolumn{6}{l}{$^{\rm c}$[S/H]~=~$-$0.81 in the DLA system}\\
\multicolumn{6}{l}{$^{\rm d}$Components {\bf a} and {\bf b};
[S/H]~$>$~$-$0.18}
\end{tabular}}
\label{tabave}
\end{table}
\par\noindent
The column densities of different species, integrated over, respectively, 
the DLA system at $z_{\rm abs}$~=~1.9733 and the sub-DLA system at 
$z_{\rm abs}$~=~1.96753 are given in Table~\ref{tabave}. For the sub-DLA
system, we have considered that most of the neutral hydrogen is associated 
with the two molecular components {\bf a} and {\bf b}. The column densities 
indicated in Table~\ref{tabave} are therefore the sum of the column densities 
measured in these two components. Note that the measurements are obtained 
using transitions with little saturation (see Fig.~4). We use in the following
the standard notation [X/H]~=~log[$Z$(X)]$-$log[$Z_{\odot}$(X)]
with $Z$(X) the metallicity of species X. Solar metallicities are from
Savage \& Sembach (1996).
\par\noindent
Sulphur and zinc are not much depleted onto dust-grains in the ISM of our 
Galaxy and therefore are good indicators of metallicity. They have consistent 
metallicities, about 0.15 solar in the DLA system and 0.7 solar in the 
sub-DLA system. 
Phosphorus is slightly enhanced ($\sim 1.5$ times) compared to Sulphur and 
Zinc in both cases. Iron, Chromium, Nickel and Manganese are underabundant 
with respect to Zinc probably because they are depleted onto dust-grains. 
\par\noindent
If we compare the measurements to what is seen in our Galaxy,
it is interesting to note that the abundance pattern seen in the 
DLA system is close to that of warm halo gas whereas the pattern in the 
sub-DLA system is close to that of warm neutral gas in the disk (see however
next Section). This is consistent with the conclusion by Petitjean et 
al. (2000) that the bulk of the DLA population is drawn from warm gas. 
\subsection{Dust depletion in the $z_{\rm abs}$~=~1.96822 component}
\begin{table}
\caption{Column densities and metallicities relative to solar
in the $z_{\rm abs}$~=~1.96822 component}
\centerline{
\begin{tabular}{lccc}
\hline
\multicolumn{1}{c}{Species}&\multicolumn{1}{c}{$N$}&
\multicolumn{1}{c}{[X/S]$_{\rm }$}&
\multicolumn{1}{c}{[X/S]$_{\rm }$}\\
\multicolumn{1}{c}{ }&\multicolumn{1}{c}{(cm$^{-2}$)}&
\multicolumn{1}{c}{Component}&\multicolumn{1}{c}{Cold ISM}\\
\hline
Mg~{\sc i}   &  3.58$\pm0.20\times10^{11}$& .... &....\\
Mg~{\sc ii}  &  1.24$\pm0.10\times10^{13}$&$-$1.33&$-$1.2\\ 
Fe~{\sc ii}  &  4.07$\pm0.23\times10^{12}$&$-$1.74&$-$2.2\\
Zn~{\sc ii}  &  2.23$\pm0.10\times10^{11}$&$-$0.15&$-$0.4\\
Si~{\sc ii}  &  3.51$\pm0.38\times10^{13}$&$-$0.85&$-$1.3\\
S~{\sc ii}   &  1.30$\pm0.10\times10^{14}$&$+$0.00&0.0\\
O~{\sc i}    &  $\ge6.0\times10^{14}$& ....&.... \\
N~{\sc i}    &  $\ge6.4\times10^{14}$& ....&....\\
P~{\sc ii}   &  $\le3.0\times10^{12}$& $<$0.06 &.... \\
Al~{\sc ii}  &  3.93$\pm0.57\times10^{11}$&$-$1.73&-2.4\\
\hline
\end{tabular}}
\label{tabmet}
\end{table}
\par\noindent
In contrast to what is seen for the system as a whole in which absorption
by Fe~{\sc ii} is strong compared to absorption by Zn~{\sc ii}, it is 
stricking to note that in the \zabs~=~1.96822 component (H$_2$ component 
{\bf b}), the absorption in the strongest iron transition 
Fe~{\sc ii}$\lambda$2382 has about the same strength as that in 
Zn~{\sc ii}$\lambda$2026 (see Fig.~\ref{metfit}). In fact, all 
transitions from refractory elements are weak. Note that Si~{\sc iii} 
absorption is clearly absent (see Fig.~\ref{metfit1}), indicating negligible 
ionization correction.
The results of Voigt profile fits to the available metal lines are given in 
Table~\ref{tabmet}. Abundances of Zn and S are similar. As can be seen 
from Table~\ref{tabmet}, Fe and Al are depleted compared to Zn by about
two orders 
of magnitude; Mg and Si are depleted by an order of magnitude compared to S. 
This is the first time that such a large depletion, consistent with that 
observed in the cold interstellar medium of the Galactic disk, is observed 
in a DLA system.
\par\noindent
\begin{figure}
\centerline{\vbox{
\psfig{figure=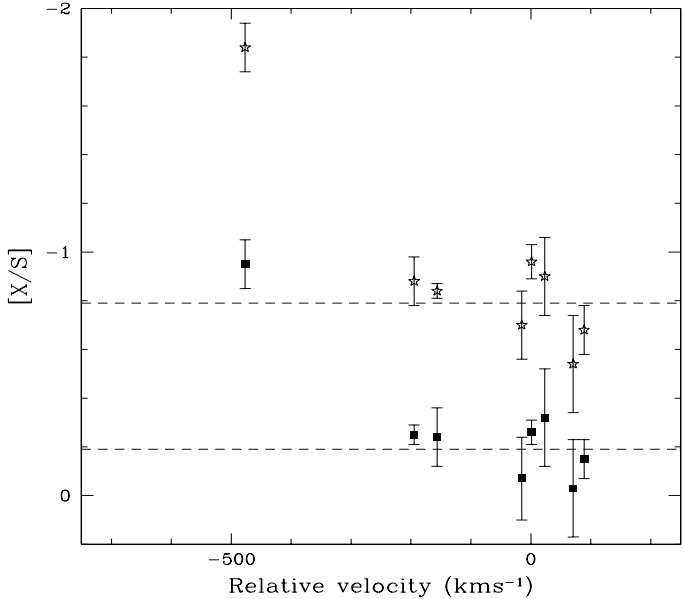,height=8.5cm,width=9.5cm,angle=0}
}}
\caption[]{Depletion of Fe (stars) and Si (squares) with respect
to S observed in different velocity components. The zero of the
velocity scale is at $z_{\rm abs}$~=~1.97296. The values averaged over
the components are indicated by horizontal dashed lines.
}
\label{depcom}
\end{figure}
\par\noindent
\subsection{Variations of dust-depletion within the system}
We investigate here the variations of the abundance ratios among different 
well separated velocity components where ionization correction is negligible. 
This is the case for the components where H$_2$ is detected. However, heavy 
element absorption from component {\bf a} is blended with absorption due to 
a nearby high-ionization feature (see Fig~\ref{metpro}). Thus we do not 
consider this component for the analysis. In addition, we consider the two 
components at $\sim-153$ and $-$195 \kms~ that account for 20 to 30\% of the 
observed heavy element column densities (see Fig.~4).
\par\noindent
We note that the [Zn/S] ratio is very close to solar (uncertainties are
typically of the order of 0.1 dex) in all components. Si and Fe abundances 
with respect to S are plotted in Fig.~\ref{depcom} for different velocity 
components. Depletion of Si and Fe is apparent. All components, except the 
very special component {\bf b} (at $-$480~km~s$^{-1}$ see previous Section), 
have similar abundance ratios (and therefore depletion factors), irrespective 
of whether \h2 molecules are present or not. This has already been 
noted  in the $z_{\rm abs}$~=~1.962 DLA system toward Q~0551$-$366 (Ledoux et 
al. 2001b).
\par
\begin{figure}
\centerline{\vbox{
\psfig{figure=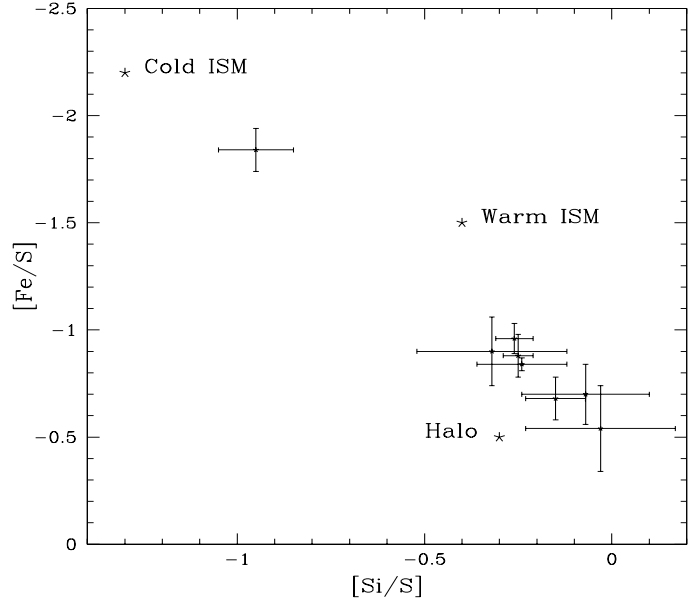,height=8.5cm,width=9.5cm,angle=0}
}}
\caption[]{[Fe/S] vs [Si/S] for different velocity components. 
The corresponding values measured in different gas components in our Galaxy 
are indicated by stars.
}
\label{cordep}
\end{figure}

\noindent
In Fig.~\ref{cordep}, [Fe/S] is plotted as a function of [Si/S] as 
observed in the sub-components. It is apparent from the figure that
there is a strong correlation between the two quantities. Note that the 
correlation is present (albeit with lower significance level) even 
if we do not consider the \zabs~=~1.96822 component. This clearly 
demonstrates the existence of differential dust-depletion in a single DLA 
system. The depletion factors for three gaseous components of our Galaxy
are overplotted on Fig.~\ref{cordep}: cold and warm ISM and halo gas
(Welty et al. 1999). It is clear that most of the velocity components have 
depletion patterns very close to that seen in warm and halo gas in our Galaxy.
Only the $z_{\rm abs}$~=~1.96822 component has a pattern similar to that of 
the cold ISM phase.
\par\noindent
It is important to note that the lowest depletion factor is recorded in 
component {\bf d} (at $+$85~km~s$^{-1}$) that has high \h2 content. In
addition, components with similar dust-depletion factors can have very 
different \h2 contents. This means that the presence of dust, though important
to form \h2 molecules at low temperature, is only one of the parameters, 
together with high particle density, low temperature and low UV-flux, that 
decide what is the molecular fraction in the gas.
\section{Physical conditions in the gas}
\begin{center}
\begin{table*}
\caption{Results of Voigt profile fitting to the C~{\sc i} absorption lines}
\begin{tabular}{lccccccc}
\hline
\multicolumn{1}{c}{$z_{\rm abs}$} &\multicolumn {1}{c}{$N$(C~{\sc i})}&
\multicolumn {1}{c}{$N$(C~{\sc i$^*$})}&
\multicolumn {1}{c}{$N$(C~{\sc i$^{**}$})$^{a}$}&
\multicolumn{1}{c}{$b$}& 
\multicolumn {1}{c}{$N$(S~{\sc ii})}&
\multicolumn{1}{c}{$T_{\rm CMBR}$}&
\multicolumn{1}{c}{$n_{\rm H}$}\\
\multicolumn{1}{c}{ } &\multicolumn {1}{c}{(10$^{12}~{\rm cm}^{-2}$)}&
\multicolumn {1}{c}{(10$^{12}~{\rm cm}^{-2}$)}&
\multicolumn {1}{c}{(10$^{12}~{\rm cm}^{-2}$)}&
\multicolumn{1}{c}{(km~s$^{-1}$)}&
\multicolumn {1}{c}{(10$^{14}~{\rm cm}^{-2}$)}&
\multicolumn{1}{c}{(K)} &
\multicolumn{1}{c}{(${\rm cm}^{-3})$} \\
\hline
\\
\multicolumn{2}{c}{ }&\multicolumn{5}{l}{Component {\bf a}
$-$ $N$(\h2)=2.39$\pm0.22\times10^{16}$~cm$^{-2}$}\\
1.96679 & 11.6$\pm$1.00&5.23$\pm$0.58& .... & 3.10$\pm$0.57 &
1.87$\pm$0.20 &$\leq$13.8 & 20-70\\
1.96691 & 7.29$\pm$0.45&3.56$\pm$0.59& .... & 3.01$\pm$0.20 &
....& $\le$14.5&20-80\\
1.96706 & 2.06$\pm$0.34&1.80$\pm$0.45& .... & 4.93$\pm$1.44 &
.... & .... & .... \\
\\
\multicolumn{2}{c}{ }&\multicolumn{5}{l}{Component {\bf b}  
$-$ $N$(\h2)=3.46$\pm0.40\times10^{16}$~cm$^{-2}$}\\
1.96822  & 10.9$\pm$0.03&12.6$\pm$0.06&5.83$\pm$0.30&4.30$\pm$0.20&
1.44$\pm$0.06 & $\le$20.0& 170-200\\
\\
\multicolumn{2}{c}{ }&\multicolumn{5}{l}{Component {\bf c} 
$-$ $N$(\h2)=0.03$-6\times10^{19}$~cm$^{-2}$}\\
1.97280 & 11.9$\pm$0.03&5.89$\pm$0.35&....&5.98$\pm$0.28&
1.42$\pm$0.18 & $\leq$13.0&40-60\\
1.97296 & 27.3$\pm$0.05&16.0$\pm$0.05&3.39$\pm$0.32&6.23$\pm$0.16&
5.68$\pm$0.34 & $\leq$13.5&50-65\\
1.97316 &4.56$\pm$0.34&2.13$\pm$0.46&....&7.00$\pm$0.77 &
0.82$\pm$0.10 & $\leq$15.0 & 10-85\\
\\
\multicolumn{2}{c}{ }&\multicolumn{5}{l}{Component {\bf d} 
$-$ $N$(\h2)=0.02$-4\times10^{19}$~cm$^{-2}$}\\
1.97365 &2.10$\pm$0.24 &$\leq$1.30   & .... & 5.66$\pm$0.41 &
0.73$\pm$0.13 & $\leq$15.0 & $\leq85$\\
1.97382 &2.52$\pm$0.23 & 1.10$\pm$0.31 & .... & 4.52$\pm$0.42 &
1.32$\pm$0.15 & $\leq$15.5 &$\leq95$\\
1.97399 & 0.81$\pm$0.23 & .... & .... & 7.30$\pm$1.02 &
....&....&....\\
1.97417 & 0.31$\pm$0.16&.... &.... & 0.37$\pm$0.27 & 
....&....&....\\
\\
\multicolumn{2}{c}{ }&\multicolumn{5}{l}{Components with
$N$(H$_2$)$\leq10^{14}{\rm cm}^{-2}$}\\
1.96737 & 8.15$\pm$0.63& 5.91$\pm$0.79 & ....& 14.41$\pm$1.08 &
1.72$\pm$0.19 & $\leq17.5$ & 40-135 \\
1.96763 &7.08$\pm$0.83& 4.76$\pm$0.99&....& 27.14$\pm$3.38 & 
1.33$\pm0.20$ & $\leq19.3$&20-180\\
1.97109 &1.90$\pm$0.12&$\leq$1.30 & .... & 10.05$\pm$0.91 &
..... & $\leq21.0$ & $\leq200$\\
1.97144 & 9.10$\pm$0.62&8.05$\pm$0.60&....& 13.44$\pm$0.96 &
.... & $\leq$18.5&65-160\\
\hline
\multicolumn{8}{l}{$^{a}$ The 2$\sigma$ upper limit is 
1.5$\times$10$^{12}$~cm$^{-2}$.}
\end{tabular}
\label{tab1}
\end{table*}
\end{center}
C~{\sc i} is usually seen in components where \h2 is detected (see however 
Srianand \& Petitjean 1998) and it is possible to derive constraints on the 
particle density of the gas from the relative populations of the different 
levels of the C~{\sc i} ground term.
\par\noindent
\begin{figure}
\centerline{\vbox{
\psfig{figure=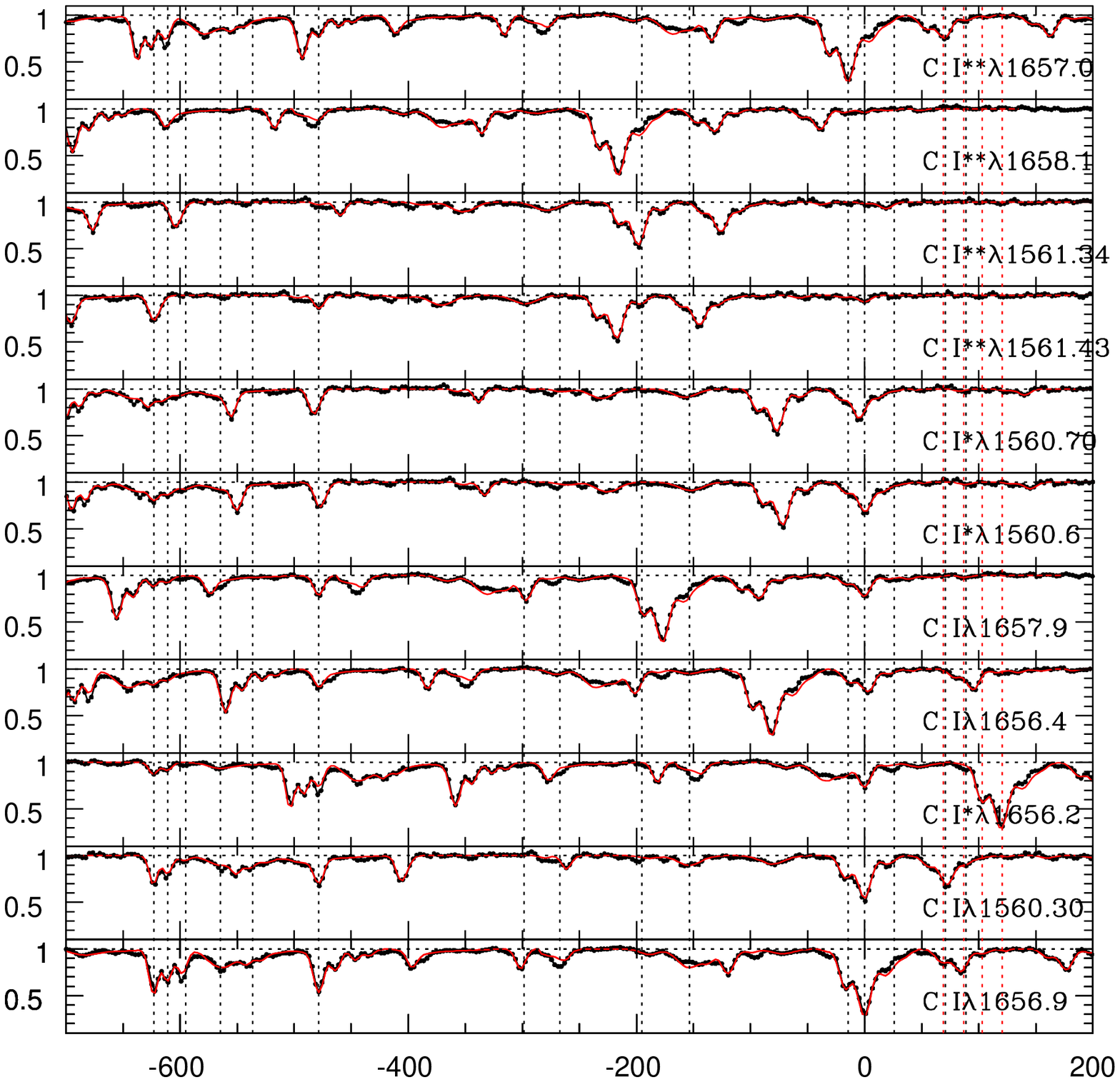,height=12.cm,width=9.5cm,angle=0}
}}
\caption[]{Velocity plot showing the absorption profiles in transitions
from different levels of the C~{\sc i} ground-term. The continuous curve gives
the best fitted profile for multiple components. The zero of the 
velocity scale is at $z_{\rm abs}$~=~1.97296. 
}
\label{cifig}
\end{figure}
Absorption lines produced by C~{\sc i} and C~{\sc i$^*$} are clearly 
seen in the four components of the DLA system toward Q\,0013$-$004
in which \h2 is detected. However, C~{\sc i$^{**}$} 
is seen only in two components at \zabs~=~1.96822 and 1.97296. In addition, 
there are three components that show C~{\sc i} absorption without detectable 
\h2 ($\le10^{14}{\rm cm}^{-2}$, see also Srianand \& Petitjean 2001). We 
perform simultaneous Voigt profile fits to the $\lambda_{\rm r}$~=~1560~ and 
1656~\AA~ C~{\sc i}, C~{\sc i$^*$} and C~{\sc i$^{**}$} lines. The fit results
are shown in Fig~\ref{cifig} and summarized in Table~\ref{tab1}. 
\par\noindent
Upper limits on the Cosmic Microwave Background Radiation (hereafter CMBR) 
temperature, $T_{\rm CMBR}$, are derived assuming 
that CMBR pumping is the only excitation mechanism. Results, given in column 
\#7 of Table~\ref{tab1}, are consistently larger than the value expected from 
standard hot big-bang model ($\sim 8$~K, see Srianand et al. 2000). This is
because other excitation processes are at play and in particular
excitation by collisions.
We estimate the particle density in the cloud assuming, $T_{\rm CMBR}$ = 8 K, 
$T_{\rm kin}$~=~100~K and neglecting fluorescence. The range given in column 
\#8 of Table~\ref{tab1} are obtained using 2$\sigma$ uncertainties 
in the column density estimate.
\par\noindent
As the ionization potential of C~{\sc i} is similar to the energy of photons
in the molecular bands, it is expected that C~{\sc i} and \h2 follow each
other. It is therefore interesting to note that $N$(C~{\sc i}) in component
{\bf d} is about a factor of 2 smaller than in component {\bf b} whereas
$N$(H$_2$) is at least a factor of ten larger. This is probably a consequence
of much larger $N$(H$_2$)/$N$(H~{\sc i}) ratio in component {\bf d}. This
conclusion is favored by the indication that the H~{\sc i} column density
is probably of the same order of magnitude in both components as indicated 
by the S~{\sc ii} column densities (see Table~\ref{tab1}).
\par\noindent
C~{\sc i}, C~{\sc i}$^*$ and C~{\sc i}$^{**}$ are clearly detected 
in component {\bf b} (see Fig~\ref{metfit}). The excitation temperature
of the $J$~=~1 \h2 level is in the range 60-80~K. Assuming that the kinetic
temperature has the same value and $T_{\rm CMBR}$~=~8~K, we derive that the 
particle density is in the range $n_{\rm H}$~=~170-200~cm$^{-3}$.
\par\noindent
It is interesting to note that the components at \zabs~=~1.96737,
and 1.96763 have a large pressure. Considering the kinetic temperature to
be 100 K implies $P/k$ = $4000-13500$ and $2000-18000$ cm$^{-3}$K respectively.
However, these components belong to the strong saturated absorption
component at $\sim$$-$600~km~s$^{-1}$ (see Fig.~2) with large 
O~{\sc i} and Si~{\sc iii} absorptions and no \h2.  It is most likely that the 
temperature of the gas is higher than what we use here (see Petitjean et al. 
2000). For $T$~=~1000 K, we derive $n$~=~10$-$75~cm$^{-3}$ and
28$-$54~cm$^{-3}$, and $P/k$~=~$18,000-61,000$ and $10,000-75,000~
{\rm cm^{-3} K}$ respectively for the two components.
\par\noindent
Such high pressures are 
seen in about one third of the C~{\sc i} gas in our galaxy (see Jenkins \& 
Tripp 2001). This probably indicates that part of the gas  in the system under
consideration is subject to high compression due to either collapse, merging
and/or supernovae explosions. The overall velocity structure of the system at 
$z_{\rm abs}$~=~1.9676 with the presence of molecular gas at $+100$ and 
$-150$~km~s$^{-1}$ from the center of the high-ionization gas strongly 
suggests the expansion of a spherical shell. In any case, intense 
star-formation activity is occuring in the vicinity of the gas.
\par\noindent
%
%
%
\section{Conclusion}
The DLA system at $z_{\rm abs}$~=~1.973 toward Q~0013$-$004 is peculiar
in several aspects. Absorption lines from metal species are spread over about 
1000~km~s$^{-1}$; in particular O~{\sc i} and C~{\sc ii} span 1050~km~s$^{-1}$.
The velocity structure indicates the presence of two main sub-systems 
centered at $z$~$\sim$~1.973 and 1.9674, separated by $\sim$~550~km~s$^{-1}$ 
with, respectively, log~$N$(H~{\sc i})~=~20.83 and $<$19.4 and 
[S/H]~=~$-$0.81 and $>-$0.18. 
\par\noindent
The low-ionization gas is conspicuous in the system. There is clear evidence 
that all species are at most twice ionized in the $z_{\rm abs}$~=~1.96822
component. This means that if photo-ionization dominates, there is probably 
very few photons with energy larger than 20~eV. More generally,
$N$(X$^{+}$)/$N$(X$^{2+}$)~$>$~1 over the entire $z$~$\sim$~1.973 system.
This ionization state could reveal gas ionized by slow shocks. This idea is 
reinforced by the high pressure measured in a few components from C~{\sc i} 
absorptions.
\par\noindent
\h2 is detected in four main components, two very strong components 
(log~$N$(\h2)~$>$~17) in the $z$~$\sim$~1.973 system and two weaker
components (log~$N$(\h2)~$\sim$~16) in the $z$~$\sim$~1.9674 system.
The total column density is 17.8~$<$~log~$N$(\h2)~$<$~20.0 and therefore the
mean molecular fraction, $f$~=~2$N$(\h2)/(2$N$(\h2)~+~$N$(H~{\sc i})), 
is in the range $-$2.7~$<$~log~$f$~$<$~$-$0.6 which is the highest molecular 
fraction observed in DLA systems. 
\par\noindent
The analysis of the $N$(C~{\sc i}$^*$)/$N$(C~{\sc i}) ratio in different
components indicates that whenever \h2 is detected, the particle density
is high ($n_{\rm H}$~$>$~30~cm$^{-3}$). High density is also found for
components without any H$_2$ absorption detected. There is a hint 
for the depletion of metals within the components to be correlated to the 
$N$(C~{\sc i}$^*$)/$N$(C~{\sc i}) ratio. This suggests that depletion
onto dust-grains could be larger for denser gas.
\par\noindent
The component at $z_{\rm abs}$~=~1.96822 shows evidence for large depletion 
of iron and silicon relative to sulfur and zinc 
([Fe/Zn]~=~$-$1.59, [Fe/S]~=~$-$1.74, [Zn/S]~=~$-$0.15, [Si/S]~=~$-$0.85)
similar to what is 
observed in cold gas of the Galactic disk. The corresponding dust-extinction 
is small in this case because, although H$_2$ is detected with 
log~$N$(H$_2$)~$\sim$~16.5, the H~{\sc i} column density is small, 
log~$N$(H~{\sc i})~$<$~19.4, in the component.  
This is direct evidence for a considerable fraction of heavy elements 
being locked into dust-grains, and, as a consequence, this supports the
idea that the current sample of DLA systems might be biased against the 
presence of cold and dusty gas along the line of sight.
Note that the rest of the gas shows a depletion pattern close to that of
warm gas in the Galactic halo.
\par\noindent
The overall kinematics of the system with the two strong sub-systems
separated by 550~km~s$^{-1}$ suggests that the line of sight is passing 
through one or several objects in strong interaction. The velocity structure 
of the subsystem at $z_{\rm abs}$~=~1.9676 with the presence of molecular 
gas at $+100$ and $-150$~km~s$^{-1}$ from the center of the high-ionization 
absorptions strongly suggests the expansion of a spherical shell.
All this, together with the strong inhomogeneity of the UV field, the high 
pressure in the gas and the high metallicities strongly suggests that
intense star-formation activity is occuring in the vicinity of the system
which should be revealed by deep imaging in the field.
%
\section*{acknowledgments}
We gratefully acknowledge support from the Indo-French Centre for 
the Promotion of Advanced Research (Centre Franco-Indien pour la Promotion
de la Recherche Avanc\'ee) under contract No. 1710-1. This work 
was supported in part by the European Communities RTN network
"The Physics of the Intergalactic Medium". RS thanks the Institute of
Astronomy in Cambridge and the Institut d'Astrophysique de Paris
and PPJ thanks IUCAA for hospitality during the time part of this work
was completed. CL acknowledges support from an ESO post-doctoral fellowship.
%


\begin{thebibliography}{}
\bibitem{}
Abgrall, H., Le Bourlot, J., Pineau des For\^ets, G., et al. 1992, A\&A, 
253, 525
\bibitem{}
Boiss\'e, P., Le Brun, V., Bergeron, J., \& Deharveng, J. M. 1998, A\&A, 333,
841
\bibitem{}
Centuri\'on, M., Bonifacio, P., Molaro, P., \& Vladilo, G. 2000, ApJ, 536, 540
\bibitem{}
D'Odorico, S., Cristiani, S., Dekker, H., et al. 2000, Proc. SPIE Vol. 4005, 
p. 121
\bibitem{}
Ellison, S. L., Yan, L., Hook, I. M., Pettini, M., Wall, J. V., \& Shaver,
P. 2001, A\&A 379, 393
\bibitem{}
Federman, S. R., Glassgold, A. E., Jenkins, E. B., \& Shaya, E. J. 1980,
ApJ, 242, 545
\bibitem{}
Ge, J., \& Bechtold, J. 1997, ApJ, 477, L73
\bibitem{}
Ge, J., Bechtold, J., \& Black, J. H. 1997, ApJ, 474, 67
\bibitem{}
Haehnelt, M. G., Steinmetz, M., \& Rauch, M. 1998, ApJ, 495, 647
\bibitem{}
Hou, J. L., Boissier, S., \& Prantzos, N. 2001, A\&A 370, 23
\bibitem{}
Jenkins, E. B., \& Tripp, T. M. 2001, ApJS, 137, 297
\bibitem{}
Ledoux, C., Bergeron, J., \& Petitjean, P. 2001a, submitted
\bibitem{}
Ledoux, C., Srianand, R., \& Petitjean, P. 2001b, submitted
\bibitem{}
Ledoux, C., Petitjean, P., Bergeron, J., Wampler, E. J., \& Srianand, R. 
1998, A\&A, 337, 51
\bibitem{}
Lu, L., Sargent, W. L. W., Barlow, T. A., Churchill, C. W., \& Vogt, S. S. 
1996, ApJS, 107, 475 
\bibitem{}
Morton, D.M. 1975, ApJ, 197, 85
\bibitem{}
Pei, Y. C., Fall, S. M., \& Bechtold, J. 1991, ApJ, 378, 6
\bibitem{}
Petitjean, P., Srianand, R., \& Ledoux, C. 2000, A\&A 364, L26
\bibitem{}
Pettini, M., Smith, L.J., King, D. L., \& Hunstead, R. W. 1997, ApJ, 486, 665
\bibitem{}
Prochaska, J. X., \& Wolfe, A. M. 1997, ApJ, 474, 140 
\bibitem{}
Prochaska, J. X., \& Wolfe, A. M. 1999, ApJS, 121, 369 
\bibitem{}
Savage, B. D., Bohlin, R. C., Drake, J. F., \& Budich, W. 1977, ApJ, 216, 291
\bibitem{}
Savage, B. D., \& Sembach, K. R., 1996, ARA\&A, 34, 279
\bibitem{}
Srianand, R., \& Petitjean, P. 1998, A\&A, 335, 33
\bibitem{}
Srianand, R., \& Petitjean, P. 2001, A\&A, 373, 816
\bibitem{}
Srianand, R., Petitjean, P., \& Ledoux, C. 2000, Nature, 408, 931
\bibitem{}
Varshalovich, D., Ivanchik, A., Petitjean, P., Srianand, R., \& Ledoux, C.
2001, AstL, 27, 683
\bibitem{}
Vladilo, G. 1998, ApJ, 493, 583
\bibitem{}
Welty, D. E., Hobbs, L. M., Lauroesch, J. T., Morton, D. C., Spitzer, L., 
York, D. C., 1999, ApJS, 124, 465
\end{thebibliography}
\end{document}